\documentclass[a4paper,11pt]{aastex631}

\usepackage{amsmath, amssymb,amsthm,amsxtra, multirow}
\usepackage[utf8]{inputenc}

\renewcommand{\baselinestretch}{1.15}
\def\thefootnote{\fnsymbol{footnote}}

\begin{document}
\title{Model Independent Approach of the JUNO $^8$B Solar Neutrino Program}

\correspondingauthor{Yufeng Li}
\email{liyufeng@ihep.ac.cn}

\correspondingauthor{Jiajie Ling}
\email{lingjj5@mail.sysu.edu.cn}

\affiliation{Yerevan Physics Institute, Yerevan, Armenia}
\affiliation{Universit\'{e} Libre de Bruxelles, Brussels, Belgium}
\affiliation{Universidade Estadual de Londrina, Londrina, Brazil}
\affiliation{Pontificia Universidade Catolica do Rio de Janeiro, Rio de Janeiro, Brazil}
\affiliation{Millennium Institute for SubAtomic Physics at the High-energy Frontier (SAPHIR), ANID, Chile}
\affiliation{Pontificia Universidad Cat\'{o}lica de Chile, Santiago, Chile}
\affiliation{Universidad Tecnica Federico Santa Maria, Valparaiso, Chile}
\affiliation{Beijing Institute of Spacecraft Environment Engineering, Beijing, China}
\affiliation{Beijing Normal University, Beijing, China}
\affiliation{China Institute of Atomic Energy, Beijing, China}
\affiliation{Institute of High Energy Physics, Beijing, China}
\affiliation{North China Electric Power University, Beijing, China}
\affiliation{School of Physics, Peking University, Beijing, China}
\affiliation{Tsinghua University, Beijing, China}
\affiliation{University of Chinese Academy of Sciences, Beijing, China}
\affiliation{Jilin University, Changchun, China}
\affiliation{College of Electronic Science and Engineering, National University of Defense Technology, Changsha, China}
\affiliation{Chongqing University, Chongqing, China}
\affiliation{Dongguan University of Technology, Dongguan, China}
\affiliation{Jinan University, Guangzhou, China}
\affiliation{Sun Yat-Sen University, Guangzhou, China}
\affiliation{Harbin Institute of Technology, Harbin, China}
\affiliation{University of Science and Technology of China, Hefei, China}
\affiliation{The Radiochemistry and Nuclear Chemistry Group in University of South China, Hengyang, China}
\affiliation{Wuyi University, Jiangmen, China}
\affiliation{Shandong University, Jinan, China, and Key Laboratory of Particle Physics and Particle Irradiation of Ministry of Education, Shandong University, Qingdao, China}
\affiliation{Institute of Modern Physics, Chinese Academy of Sciences, Lanzhou, China}
\affiliation{Nanjing University, Nanjing, China}
\affiliation{Guangxi University, Nanning, China}
\affiliation{East China University of Science and Technology, Shanghai, China}
\affiliation{School of Physics and Astronomy, Shanghai Jiao Tong University, Shanghai, China}
\affiliation{Tsung-Dao Lee Institute, Shanghai Jiao Tong University, Shanghai, China}
\affiliation{Institute of Hydrogeology and Environmental Geology, Chinese Academy of Geological Sciences, Shijiazhuang, China}
\affiliation{Nankai University, Tianjin, China}
\affiliation{Wuhan University, Wuhan, China}
\affiliation{Xi'an Jiaotong University, Xi'an, China}
\affiliation{Xiamen University, Xiamen, China}
\affiliation{School of Physics and Microelectronics, Zhengzhou University, Zhengzhou, China}
\affiliation{Institute of Physics, National Yang Ming Chiao Tung University, Hsinchu}
\affiliation{National United University, Miao-Li}
\affiliation{Department of Physics, National Taiwan University, Taipei}
\affiliation{Charles University, Faculty of Mathematics and Physics, Prague, Czech Republic}
\affiliation{University of Jyvaskyla, Department of Physics, Jyvaskyla, Finland}
\affiliation{IJCLab, Universit\'{e} Paris-Saclay, CNRS/IN2P3, 91405 Orsay, France}
\affiliation{Univ. Bordeaux, CNRS, LP2i Bordeaux, UMR 5797, F-33170 Gradignan, France}
\affiliation{IPHC, Universit\'{e} de Strasbourg, CNRS/IN2P3, F-67037 Strasbourg, France}
\affiliation{Centre de Physique des Particules de Marseille, Marseille, France}
\affiliation{SUBATECH, Nantes Universit\'{e}, IMT Atlantique, CNRS-IN2P3, Nantes, France}
\affiliation{III. Physikalisches Institut B, RWTH Aachen University, Aachen, Germany}
\affiliation{Institute of Experimental Physics, University of Hamburg, Hamburg, Germany}
\affiliation{Forschungszentrum J\"{u}lich GmbH, Nuclear Physics Institute IKP-2, J\"{u}lich, Germany}
\affiliation{Institute of Physics and EC PRISMA$^+$, Johannes Gutenberg Universit\"{a}t Mainz, Mainz, Germany}
\affiliation{Technische Universit\"{a}t M\"{u}nchen, M\"{u}nchen, Germany}
\affiliation{Helmholtzzentrum f\"{u}r Schwerionenforschung, Planckstrasse 1, D-64291 Darmstadt, Germany}
\affiliation{Eberhard Karls Universit\"{a}t T\"{u}bingen, Physikalisches Institut, T\"{u}bingen, Germany}
\affiliation{INFN Catania and Dipartimento di Fisica e Astronomia dell Universit\`{a} di Catania, Catania, Italy}
\affiliation{Department of Physics and Earth Science, University of Ferrara and INFN Sezione di Ferrara, Ferrara, Italy}
\affiliation{INFN Sezione di Milano and Dipartimento di Fisica dell Universit\`{a} di Milano, Milano, Italy}
\affiliation{INFN Milano Bicocca and University of Milano Bicocca, Milano, Italy}
\affiliation{INFN Milano Bicocca and Politecnico of Milano, Milano, Italy}
\affiliation{INFN Sezione di Padova, Padova, Italy}
\affiliation{Dipartimento di Fisica e Astronomia dell'Universit\`{a} di Padova and INFN Sezione di Padova, Padova, Italy}
\affiliation{INFN Sezione di Perugia and Dipartimento di Chimica, Biologia e Biotecnologie dell'Universit\`{a} di Perugia, Perugia, Italy}
\affiliation{Laboratori Nazionali di Frascati dell'INFN, Roma, Italy}
\affiliation{University of Roma Tre and INFN Sezione Roma Tre, Roma, Italy}
\affiliation{Institute of Electronics and Computer Science, Riga, Latvia}
\affiliation{Pakistan Institute of Nuclear Science and Technology, Islamabad, Pakistan}
\affiliation{Joint Institute for Nuclear Research, Dubna, Russia}
\affiliation{Institute for Nuclear Research of the Russian Academy of Sciences, Moscow, Russia}
\affiliation{Lomonosov Moscow State University, Moscow, Russia}
\affiliation{Comenius University Bratislava, Faculty of Mathematics, Physics and Informatics, Bratislava, Slovakia}
\affiliation{Department of Physics, Faculty of Science, Chulalongkorn University, Bangkok, Thailand}
\affiliation{National Astronomical Research Institute of Thailand, Chiang Mai, Thailand}
\affiliation{Suranaree University of Technology, Nakhon Ratchasima, Thailand}
\affiliation{Department of Physics and Astronomy, University of California, Irvine, California, USA}

\author{Jie Zhao}
\affiliation{Institute of High Energy Physics, Beijing, China}

\author{Baobiao Yue}
\affiliation{Sun Yat-Sen University, Guangzhou, China}

\author{Haoqi Lu}
\affiliation{Institute of High Energy Physics, Beijing, China}

\author{Yufeng Li}
\affiliation{Institute of High Energy Physics, Beijing, China}

\author{Jiajie Ling}
\affiliation{Sun Yat-Sen University, Guangzhou, China}

\author{Zeyuan Yu}
\affiliation{Institute of High Energy Physics, Beijing, China}

\author{Angel Abusleme}
\affiliation{Pontificia Universidad Cat\'{o}lica de Chile, Santiago, Chile}
\affiliation{Millennium Institute for SubAtomic Physics at the High-energy Frontier (SAPHIR), ANID, Chile}

\author{Thomas Adam}
\affiliation{IPHC, Universit\'{e} de Strasbourg, CNRS/IN2P3, F-67037 Strasbourg, France}

\author{Shakeel Ahmad}
\affiliation{Pakistan Institute of Nuclear Science and Technology, Islamabad, Pakistan}

\author{Rizwan Ahmed}
\affiliation{Pakistan Institute of Nuclear Science and Technology, Islamabad, Pakistan}

% ...

\begin{abstract}
The physics potential of detecting $^8$B solar neutrinos {will be} exploited at the Jiangmen Underground Neutrino Observatory (JUNO), in a model independent manner by using three distinct channels of the charged-current (CC), neutral-current (NC) and elastic scattering (ES) interactions.
Due to the largest-ever mass of $^{13}$C nuclei in the liquid-scintillator detectors and the {expected} low background level, $^8$B solar neutrinos would be observable in the CC and NC interactions on $^{13}$C for the first time.
By virtue of optimized event selections and muon veto strategies, backgrounds from the accidental coincidence, muon-induced isotopes, and external backgrounds can be greatly suppressed. Excellent signal-to-background ratios can be achieved in the CC, NC and ES channels to guarantee the $^8$B solar neutrino observation.
From the sensitivity studies performed in this work, {we show that JUNO, with ten years of data}, can reach the {1$\sigma$} precision levels of 5\%, 8\% and 20\% for the $^8$B neutrino flux, $\sin^2\theta_{12}$, and $\Delta m^2_{21}$, respectively. It would be unique and helpful to probe the details of both solar physics and neutrino physics. In addition, when combined with SNO, the world-best precision of 3\% is expected for the $^8$B neutrino flux measurement. 

\end{abstract}

\section{Introduction}
%
Electron neutrino fluxes are produced from thermal nuclear fusion reactions in the solar core, either through the proton-proton ($pp$) chain or the Carbon-Nitrogen-Oxygen (CNO) cycle. According to their production reactions, the solar neutrino species can be categorized as $pp$, $^7$Be, $pep$, $^8$B, $hep$  neutrinos of the $pp$ chain, and $^{13}$N, $^{15}$O, and $^{17}$F neutrinos of the CNO cycle.
Before reaching the detector, solar neutrinos undergo the flavor conversion inside the Sun and the Earth during their propagation.
{Solar neutrino measurements have a long history starting with the measurements done by the Homestake experiment~\citep{Davis:1968cp}.}
Many measurements, such as Homestake~\citep{Davis:1968cp}, Kamiokande~\citep{Kamiokande-II:1989hkh}, GALLEX/GNO~\citep{GALLEX:1993php,GNO:2000avz}, SAGE~\citep{Abazov:1991rx}, and Super-Kamiokande (SK)~\citep{Super-Kamiokande:1998qwk,Super-Kamiokande:2001ljr}, had observed the solar neutrino deficit problem: that is the amount of observed neutrinos originating from the Sun was much less than that expected from the Standard Solar Model (SSM). Subsequently, the Sudbury Neutrino Observatory (SNO) provided the first model-independent evidence of solar neutrino flavor conversion using three distinct neutrino interaction channels in heavy water~\citep{Chen:1985na,SNO:2001kpb,SNO:2002tuh,SNO:2003bmh,SNO:2008gqy,SNO:2011ajh,SNO:2011hxd}. These reactions {are} the $\nu_e$ sensitive charged-current (CC) interaction, all flavor sensitive neutral-current (NC) interaction on Deuterium, and the elastic scattering (ES) interaction on electrons from all neutrino flavors with different cross sections.

Solar neutrino observations {rely on the SSM flux predictions, the neutrino oscillation parameters and solar density model} that determine the flavor conversion~\citep{Wolfenstein:1978ue,Mikheyev:1985zog,ParticleDataGroup:2020ssz}. 
Thus although SK~\citep{Super-Kamiokande:2016yck,Super-Kamiokande:2013mie} and Borexino~\citep{BOREXINO:2014pcl,BOREXINO:2020aww} experiments have made precision measurements on the $^8$B neutrinos via the ES interaction, the evaluation of the total amount of neutrinos produced inside the Sun relies on the input of solar neutrino oscillations~\citep{ParticleDataGroup:2020ssz}.
The present most precise $^8$B neutrino flux is determined by SNO with the {1$\sigma$ confidence level uncertainly of around} 3.8\%~\citep{SNO:2002tuh,SNO:2003bmh,SNO:2008gqy,SNO:2011ajh,SNO:2011hxd}, and it is the only existing model independent flux measurement. Therefore, a second independent measurement of the total $^8$B neutrino flux with the NC channel~\citep{Arafune:1988hx,Ianni:2005ki} would be important to answer relevant questions in the field of solar physics. For example, there is the solar abundance problem, in which the SSM based on the solar composition with a higher value of metallicity is inconsistent with the helioseismological measurements~\citep{Vinyoles:2016djt}. {Note that a recent solar model is able to resolve the discord between the helioseismological and photospheric measurements~\citep{Magg:2022rxb}, but lively discussions on this topic are still on-going~\citep{Buldgen:2022nso,Yang:2022dsq}.}

In contrast, the neutrino oscillation parameters $\sin^2\theta_{12}$ and $\Delta m^2_{21}$ have reached the {1$\sigma$ confidence level uncertainty of} around 5\% and 15\% respectively, {from the current global solar neutrino data~\citep{Esteban:2020cvm}.} The mixing angle $\sin^2\theta_{12}$ is extracted from the comparison of the observed fluxes of $pp$, $^7$Be, and $^8$B solar neutrinos to their respective total fluxes from the SSM. And the mass squared difference $\Delta m^2_{21}$ is measured from both the vacuum-matter transition of the $^8$B neutrino oscillations and the size of the day-night asymmetry. A direct comparison of oscillation parameters from the solar neutrino and reactor antineutrino oscillations is an unique probe of new physics beyond the Standard Model of particle physics. It would be excellent to have a new measurement of solar neutrino oscillations with high precision in this respect. This has triggered a variety of interesting discussions on the prospects of future large neutrino detectors~\citep{Capozzi:2018dat,JUNO:2020hqc,Hyper-Kamiokande:2018ofw,Jinping:2016iiq}.


The Jianmen Underground Neutrino Observatory (JUNO) is a liquid scintillator (LS) detector of 20 kton, which is located in South China and will start data taking by 2024. As a multiple-purpose neutrino experiment, JUNO is unique for the solar neutrino detection because of its large target mass, excellent energy resolution, and expected low background levels. 
With the analysis threshold cut of around 2 MeV for the recoiled electron energies in the ES channel, JUNO can make a high-statistics measurement of the flux and spectral shape of $^8$B solar neutrinos and will be able to extract the neutrino oscillation parameters $\sin^2\theta_{12}$ and $\Delta m^2_{21}$~\citep{JUNO:2020hqc}.
In addition to the high statistics measurement in the ES channel, the presence of a large mass of the $^{13}$C nuclei ($\sim$0.2 kt) makes it feasible to detect $^8$B solar neutrinos via CC and NC interactions on $^{13}$C. 
By combining the CC, NC and ES channels, we are able to perform a model independent measurement of the $^8$B solar neutrino flux and oscillation parameters $\sin^2\theta_{12}$ and $\Delta m^2_{21}$, which will add a unique contribution to the global solar neutrino program.

The paper is organized as follows.
We illustrate the typical signatures of the CC and NC interactions of $^8$B solar neutrinos, and evaluate the corresponding backgrounds in the JUNO detector in Sec.~\ref{sec2}. In Sec.~\ref{sec3},
the physics potential of detecting the $^8$B solar neutrinos with different combinations of the CC, NC, and ES channels are presented, and the sensitivity to the $^8$B solar neutrino flux, $\sin^2\theta_{12}$ and $\Delta m^2_{21}$ is reported. The concluding remarks of this study are presented in Sec.~\ref{sec4}.

\section{Signal and Background at JUNO}
\label{sec2}
%

The JUNO experiment is building the world largest LS detector with the total target mass of 20 kt, in which the mass fraction of Carbon is 88\%. Given that the natural abundance of $^{13}$C is 1.1\%, the total mass of $^{13}$C {reaches 193.6\,ton, which is similar to the total Deuterium mass of 200\,ton for the SNO detector. Considering the preferable cross sections of $^{13}$C at the solar neutrino energies~\citep{Fukugita:1988hg,Suzuki:2012aa,Suzuki:2019cra}, the CC and NC solar neutrino rates on $^{13}$C will be rather sizable in the JUNO detector.}

In Table~\ref{tab:C13signal}, we present the typical CC, NC and ES detection channels for $^8$B solar neutrinos in the LS medium. For each interaction channel, the reaction threshold is provided, together with the typical experimental signatures, and the expected event numbers for 10 years of data taking before event selection cuts.
The spin and parity of the daughter nuclei at the ground (gnd) or excited state, denoted by the corresponding excited energies, are also provided.
The unoscillated $^8$B solar neutrino $\nu_e$ flux (5.25$\times$10$^6$ /cm$^2$/s) is taken from the final result of SNO for this estimation~\citep{SNO:2011hxd}, and the {spectrum is} taken from ~\cite{Bahcall:1996qv,Bahcall:1997eg}. The cross sections for these exclusive channels are taken from the calculation in ~\cite{Fukugita:1988hg,Suzuki:2012aa,Suzuki:2019cra}, in which the uncertainties
at the level of a few percent are considered to be achievable. Note that the standard Mikheev-Smirnov-Wolfenstein (MSW) effect of solar neutrino oscillations~\citep{Wolfenstein:1978ue,Mikheyev:1985zog} and the neutrino oscillation parameters from ~\cite{ParticleDataGroup:2020ssz} are used in the signal calculations of the CC, NC, and ES channels.

\begin{table*}[tb]\small
  \begin{minipage}[c]{\textwidth}
  \caption{Typical CC, NC, and ES detection channels of the $^8$B solar neutrinos together with the final states, the neutrino energy threshold, the typical signatures in the detector, and the expected event numbers with 10 years of data taking. Note that $\nu_{x}$ with ($x=e,\mu,\tau$) denotes all three active flavor neutrinos. The spin and parity of the daughter nuclei at the ground (gnd) or excited states, denoted as the corresponding excited energies, are also provided. }
    \vspace{0.5cm}
  \resizebox{\textwidth}{!}{
	\begin{tabular}{c|c|c|c|c|c}
        \hline
    No. & \multicolumn{2}{c|} {Channels} & Threshold [MeV] & Signal & Event numbers (10 years) \\ \hline
    1 &  & $\nu_{e}+^{12}{\rm C}\rightarrow e^-+^{12}{\rm N}\,(1^+;{\rm gnd})$~\citep{Fukugita:1988hg} & 16.827 & $e^-$+$^{12}$N decay ($\beta^{+}$, Q=17.338\,MeV) & 0.43 \\
    1 & \multirow{2}{*}{CC} & $\nu_e+^{13}{\rm C}\rightarrow e^-+^{13}{\rm N}\,(\frac{1}{2}^{-};{\rm gnd})$~\citep{Suzuki:2012aa} & 2.2 & $e^-$+$^{13}$N decay ($\beta^{+}$, Q=2.22\,MeV) &  3929 \\
    2 & & $\nu_e+^{13}{\rm C}\rightarrow e^-+^{13}{\rm N}\,(\frac{3}{2}^{-};3.5\,{\rm MeV})$~\citep{Suzuki:2012aa} & 5.7 & $e^-$+$p$ &  2464 \\ \hline
    4 &  & $\nu_{x}+^{12}{\rm C}\rightarrow\nu_{x}+^{12}{\rm C}\,(1^+;15.11\,{\rm MeV})$~\citep{Fukugita:1988hg} & 15.1 & $\gamma$ & 4.8 \\
    3 & \multirow{5}{*}{NC} & $\nu_{x}+^{13}{\rm C}\rightarrow\nu_{x}+n+^{12}{\rm C}\,(2^+;4.44\,{\rm MeV})$~\citep{Suzuki:2019cra} & 6.864 & $\gamma$ + $n$ capture & 65 \\
    4 & & $\nu_{x}+^{13}{\rm C}\rightarrow\nu_{x}+^{13}{\rm C}\,(\frac{1}{2}^+;3.089\,{\rm MeV})$~\citep{Suzuki:2012aa} & 3.089 & $\gamma$ & 14 \\
    5 & & $\nu_{x}+^{13}{\rm C}\rightarrow\nu_{x}+^{13}{\rm C}\,(\frac{3}{2}^-;3.685\,{\rm MeV})$~\citep{Suzuki:2012aa} & 3.685 & $\gamma$ & 3032 \\
    6 & & $\nu_{x}+^{13}{\rm C}\rightarrow\nu_{x}+^{13}{\rm C}\,(\frac{5}{2}^+;3.854\,{\rm MeV})$~\citep{Suzuki:2012aa} & 3.854 & $\gamma$ & 2.8 \\ \hline
    7 & ES & $\nu_{x}+e\rightarrow\nu_{x}+e$ & 0 & $e^-$ & $3.0\times10^{5}$  \\
    \hline
	\end{tabular}}
	\label{tab:C13signal}
\end{minipage}
\end{table*}

There are no interactions on the $^{12}$C nuclei for most solar neutrinos because of the
high energy threshold. Thus for the CC channel, we are left
with the following two exclusive interactions:
{\begin{eqnarray}
\nu_e+^{13}{\rm C}&\rightarrow& e^-+^{13}{\rm N}\,\left(\frac{1}{2}^{-}; {\rm gnd}\right)\;,\\
\nu_e+^{13}{\rm C}&\rightarrow& e^-+^{13}{\rm N}\,\left(\frac{3}{2}^{-}; 3.502\,{\rm MeV}\right)\;,
\end{eqnarray}}
where the final $^{13}{\rm N}$ is in the ground state and {excited $^{13}{\rm N}\,({3}/{2}^-;3.502\,{\rm MeV})$ state respectively. For the first reaction channel,} the ground state of $^{13}{\rm N}$ undergoes a delayed $\beta^+$ decay ($Q$ = 2.2 MeV) with a lifetime of 863 s. The distinct signature for this channel is a coincidence of the prompt electron and delayed positron with stringent time, distance, and energy requirements. The expected number of events for $^8$B solar neutrinos in this coincidence channel is 3929 for 10 years of data taking.
{On the other hand,
although the channel with an excited $^{13}{\rm N}\,({3}/{2}^-;3.502\,{\rm MeV})$ has a comparable cross section as the ground-state channel~\citep{Suzuki:2012aa}, the corresponding signature after quenching is a single event since the deexcitation of $^{13}{\rm N}\,({3}/{2}^-;3.502\,{\rm MeV})$ is dominated by a proton knockout, and thus cannot be distinguished from the recoiled electron of the ES channel and the single $\gamma$ of the NC channel on an event-by-event basis. Therefore, in the coincidence event category we focus on the CC channel with the ground state $^{13}{\rm N}$ and consider the channel with the excited $^{13}{\rm N}\,({3}/{2}^-;3.502\,{\rm MeV})$ as a component of the total singles spectrum as illustrated in Fig.~\ref{fig:specNC-ES}.}

Among the five listed NC channels, the only one with a coincidence signature is the interaction of $\nu_{x}+^{13}{\rm C}\rightarrow\nu_{x}+n+^{12}{\rm C}$, with a prompt $\gamma$ energy of 4.44 MeV from $^{12}$C de-excitation and the delayed neutron capture. However, given that the background from the inverse beta decay interactions of reactor antineutrinos are overwhelming, where the signal to background ratio is at the level of 10$^{-4}$, and thus the event rate of this channel is unobservable. In this work we focus on the NC channels with the signature of single $\gamma$ deexcitation, among which the NC interaction with the $^{13}$C de-excited energy of 3.685 MeV:
\begin{equation}
\nu_{x}+^{13}{\rm C}\rightarrow\nu_{x}+^{13}{\rm C}\,\left(\frac{3}{2}^-;3.685\,{\rm MeV}\right)\;,
\end{equation}
is the dominant interaction channel and will be used to determine the $^8$B solar neutrino flux via the NC interaction.

Finally, we also consider the ES interaction channel on the electron,
\begin{equation}
\nu_x+e\rightarrow\nu_x+e\;,
\end{equation}
where the signature is a single recoiled electron~\citep{JUNO:2020hqc}. Using all the three channels of CC, NC, ES interactions, we are able to make a model independent measurement of the $^8$B solar neutrino flux, $\sin^2\theta_{12}$ and $\Delta m^2_{21}$ with JUNO, which is useful to disentangle the solar dynamics and the neutrino oscillation effects. This measurement is expected to be the only model independent study after the SNO experiment~\citep{SNO:2001kpb,SNO:2002tuh,SNO:2003bmh,SNO:2008gqy}.

To summarize, in this work we are going to employ the following three interaction channels for a model independent approach of the JUNO $^8$B solar neutrino program: i) the CC detection channel is sensitive to the $\nu_{e}$ component of solar neutrinos, ii) the NC channel is sensitive to all active neutrino flavors ($\nu_{e}$, $\nu_{\mu}$, $\nu_{\tau}$) with identical cross sections, iii) the ES channel is also sensitive to all active flavors, but with a preferred cross section for the $\nu_{e}$ flux [i.e., $\sigma(\nu_{\mu/\tau})\simeq0.17\,\sigma(\nu_{e})$].

\subsection{$\nu_e+^{13}{\rm C}$ Charged Current Channel}

For the typical coincidence signature of the CC channel, $\nu_e+^{13}{\rm C}\rightarrow e^-+^{13}{\rm N}\,({1}/{2}^{-};{\rm gnd})$, the energy of the prompt signal is the kinetic energy of the outgoing electron with the reaction threshold of 2.2 MeV. Therefore, there is a one-to-one correspondence between the electron kinetic energy and the initial neutrino energy $T_{e} \simeq E_{\nu} - 2.2$ MeV, because of the negligible recoil energy of the daughter $^{13}$N. Meanwhile, the delayed signal is the deposited energy of the positron from the $^{13}$N $\beta^+$ decay ($Q = 2.2$ MeV), with a decay lifetime of $\tau$ = 863 s. The time and spatial correlation between the prompt and delayed signals provides the distinct feature of the coincidence signature.

\begin{table*}
  \begin{minipage}[c]{\textwidth}
  \caption{ \label{tab:CCeff}
     The efficiencies of optimized event selection cuts for the signal and backgrounds of the $\nu_e$ CC channel [$\nu_e+^{13}{\rm C}\rightarrow e^-+^{13}{\rm N}\,({1}/{2}^{-};{\rm gnd})$] analysis. The expected event numbers of the signal and backgrounds  for 10 years of data taking after each cut are also listed. The fiducial volume used in this work corresponds to the effective mass of 16.2 kt. For the energy cuts, $E_p$ and $E_d$ represent the visible energy of prompt and delayed signals. The same muon and three-fold-coincidence veto strategies as in ~\cite{JUNO:2020hqc} are used for the reduction of muon-induced isotopes.}
         \vspace{0.5cm}
  \resizebox{\textwidth}{!}{
     
	\begin{tabular}{c|c|c|c|c|c|c}
        \hline
    & \multirow{3}{*}{Cuts} & \multirow{3}{*}{CC signal efficiency} & \multirow{3}{*}{CC signal} & \multicolumn{3}{c}{ Background for CC channel} \\ \cline{5-7}
    & & & & Solar ES & \multicolumn{2}{c}{Muon-induced isotopes} \\
    \cline{5-7}
    &&&& Accidental & Accidental & Correlated \\ \hline
    -- & -- & -- & 3929 & -- & -- & -- \\ \hline
    Time cut & 4 ms $<\Delta T <$ 900 s & 65\% & 2554 & 10$^{10}$ & 10$^{13}$ & 10$^{12}$ \\ \hline
    \multirow{2}{*}{Energy cut} &  5 MeV $< E_p < $14 MeV & 79\% & \multirow{2}{*}{1836} & \multirow{2}{*}{10$^9$} & \multirow{2}{*}{10$^{10}$} & \multirow{2}{*}{10$^9$} \\
    & 1 MeV $< E_d <$ 2 MeV & 91\% & && \\ \hline
        Fiducial volume Cut & $R<$ 16.5 m~\citep{JUNO:2020hqc} & 81\% & 1487 & 10$^7$ & 10$^7$ & 10$^8$\\
    \hline
    Vertex cut & $\Delta d <$ 0.47 m & 87\% &1293 & 328 & 10$^5$ & 10$^6$ \\
\hline
    Muon veto & Muon and TFC veto~\citep{JUNO:2020hqc} & 50\% & 647 & 164 & 53 & 58 \\ \hline
    Combined & -- & 17\% & 647 & \multicolumn{3}{c}{275} \\
    \hline
    \end{tabular}}
    \end{minipage}
\end{table*}

\begin{figure}
\begin{center}
	\includegraphics[width=0.6\textwidth]{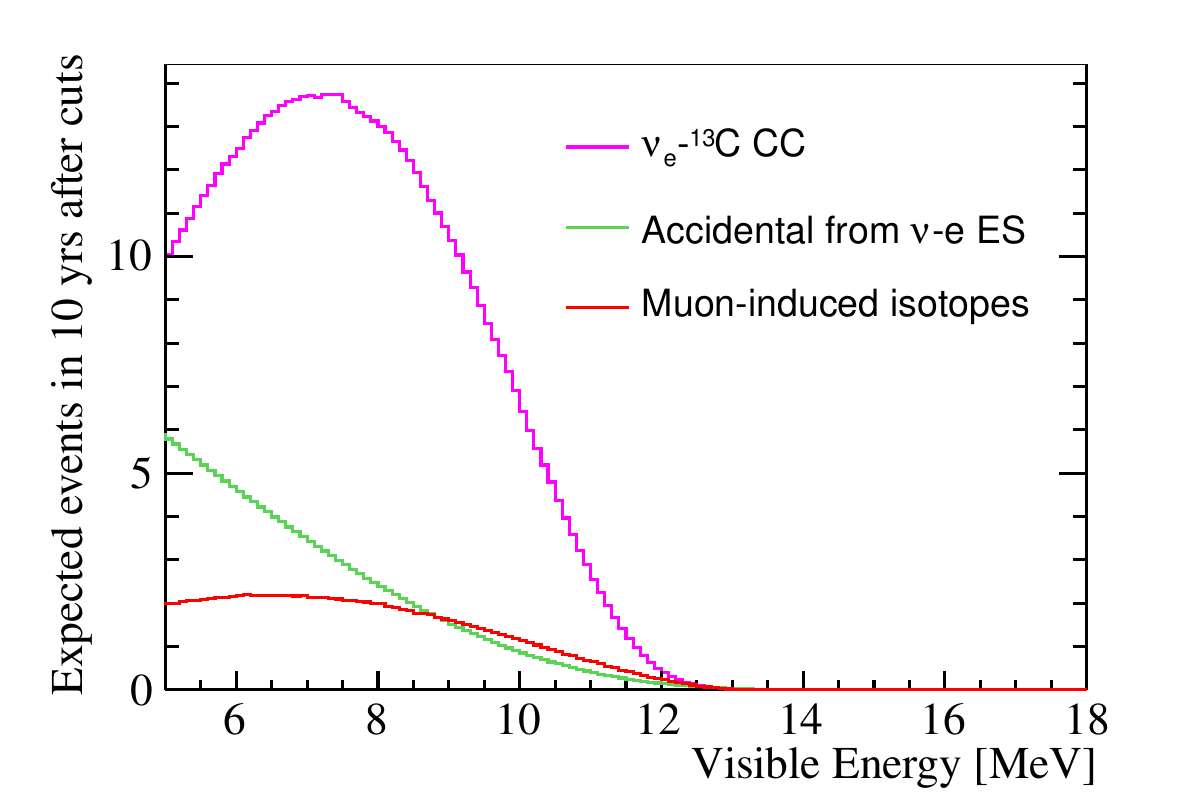}
    \caption{Expected prompt visible energy spectra of the CC signal and backgrounds after the optimized cuts. {The y-axis represents the number of events per 0.1 MeV.} The accidental background with the recoiled electron from solar neutrino ES interaction as the prompt signal is illustrated as the green line. The background from muon-induced isotopes is illustrated as the red line, which is the summation of the accidental and correlated backgrounds originated from the initial muons.}\label{fig:CCspec}
\end{center}
\end{figure}

In the following, we consider two significant backgrounds for this coincidence signature of the CC channel in this work.

\begin{itemize}
\item The first background is the accidental coincidence of two single events.
For the visible energy between 2 and 5 MeV, natural radioactivity composes the most significant part of the prompt component of the coincidence candidate, while {the prompt background events} above 5 MeV come from the muon-induced unstable isotopes and the recoiled electrons of solar neutrino ES interactions. 
Due to the expected natural radioactivity level in the LS (10$^{-17}$ g/g $^{238}$U and $^{232}$Th in the secular equilibrium, 10$^{-18}$ g/g $^{40}$K and 10$^{-24}$ g/g $^{210}$Pb), a requirement on the selection of the prompt energy is to minimize the contribution from these radioactivity events. The delayed component of the accidental background is mainly from the cosmogenic $^{11}$C decay ($Q$ = 1.98 MeV) in the energy range of [1, 2] MeV, while the internal LS radioactivity contributes less than 2\% compared to that from $^{11}$C. If the internal radioactivity is 1-2 orders of magnitude higher than expected, the contribution to the delayed component from the radioactivity would be at the same level as the cosmogenic $^{11}$C decay. Note that all the single events in the energy range between 1 and 2 MeV can be accurately measured in-situ with the future data, and the accidental background can be deduced with the off-time coincidence method. Note that we have neglected the external radioactivity which can be effectively removed by the proper fiducial volume cut.

\item The second background is produced by the correlated prompt and delayed decays of unstable isotopes from the same parent muon. These correlated decays are not considered in the above accidental background. Therefore, the cosmic muon and the corresponding isotope simulations have been performed, and the muon veto strategies of the three-fold-coincidence are the same as those in ~\cite{JUNO:2020hqc}. It shows that the prompt signal is mainly from the beta decays of $^{12}$B, $^{8}$Li, $^6$He, and  $^{10}$C (below 4 MeV), and as expected the delayed signal is from $^{11}$C.
The muon detection efficiency of the outer water veto can reach as high as 99.5\%~\citep{JUNO:2020hqc}. Since the remaining untagged muons are usually located at the edge of the central detector, these muon-induced correlated background can be removed using the fiducial volume cut and is neglected in this work. Note that we have assumed a perfect detector uniformity for these isotopes and used the whole detector region to estimate the background inside the fiducial volume.
\end{itemize}

We have simulated the signal and backgrounds using the official JUNO simulation software~\citep{Lin:2017usg,Zou:2015ioy}. 
According to the signal characteristics of the CC channel, the accidental background can be calculated with different selection cuts. The final event selection criteria is obtained by optimizing the figure of merit, $S/\sqrt{S+B}$, where $S$ and $B$ stand for the rates of the signal and background, respectively. 
The optimized event selection cuts of the fiducial volume, the prompt and delayed energies, the time and spatial correlation cuts and muon vetos are provided step by step in Tab.~\ref{tab:CCeff}, where the efficiencies of the signal and backgrounds are also calculated. In order to avoid possible large contamination from the internal radioactivity and muon-induced $^{10}$C, we select the threshold of the prompt visible energy to 5 MeV for the CC channel, i.e., 5 MeV $< E_p < $ 14 MeV. Meanwhile, the fiducial volume is chosen to be $R<$ 16.5 m to reject the external radioactivity and isotopes, with $R$ being the distance to the detector center.
{It should be noted that an anti-coincidence criterion with a time distance cut of $\Delta T>$ 4 ms has been used to reject the inverse beta decay (IBD) interactions of reactor antineutrinos, achieving a rejection power of 100\%. Meanwhile, this IBD rejection cut has negligible impact on the signal because of the much longer lifetime of $^{13}$N.}

We illustrate in Fig.~\ref{fig:CCspec} the expected prompt visible energy spectra of the selected signal and residual backgrounds in the CC channel after the optimized cuts. The expected number of selected signals is 647 for 10 years of data taking, which is shown as the purple line. 
The fiducial volume used in this work corresponds to the effective mass of 16.2 kt.
The accidental background with solar neutrino ES interactions as the prompt signal is illustrated as the green line and contributes 164 background events, which will be fully correlated with the solar neutrino ES signal in the following global analysis. In contrast, the muon-induced isotopes contribute 111 background events (depicted as the red line of Fig.~\ref{fig:CCspec}), which are from both the accidental coincidence (53 events) and correlated background (58 events). Therefore, we can achieve an excellent $S/\sqrt{S+B}$ $\simeq$ 21, {offering an excellent prospect for the future experimental measurements. As a comparison, a preliminary study assessing the feasibility of detecting solar neutrinos via the CC interactions on $^{13}$C in the Borexino experiment has been previously reported in the thesis of Chiara Ghiano~\citep{Ghiano:2012pny}, where an upper limit for the number of the solar neutrino CC interaction $^{13}$C was established, constrained by the limited event statistics.}

Finally, the expected event number of $hep$ solar neutrinos in the CC channel is about 15 for ten years of data taking, but only 3 events are beyond the spectral tail of $^8$B solar neutrinos. Thus it would be difficult to detect the $hep$ solar neutrinos with the CC interaction on $^{13}$C, and the signal from the $hep$ solar neutrinos will be neglected in this work.


\subsection{$\nu_x+^{13}{\rm C}$ Neutral Current Channel}

\begin{figure}
\begin{center}
	\includegraphics[width=0.6\textwidth]{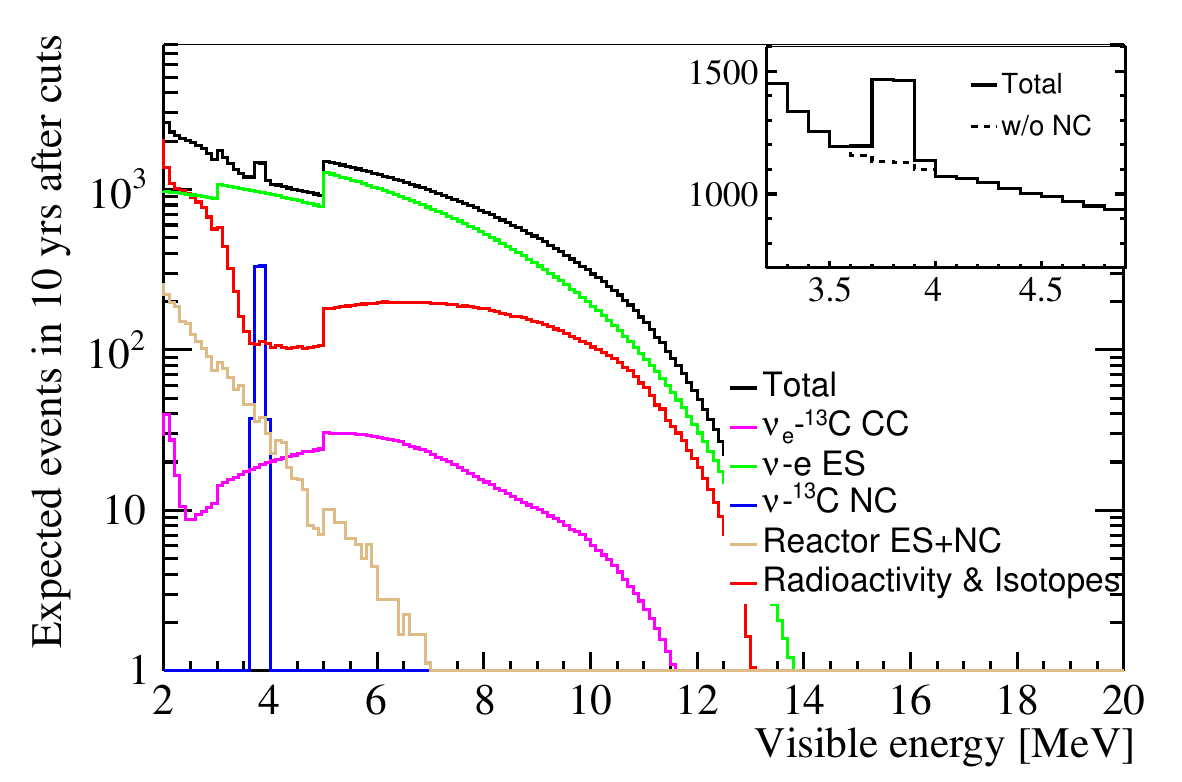}
    \caption{Expected visible energy spectra of all single event sources for 10 years of data taking with the same energy-dependent fiducial volume cuts as in ~\cite{JUNO:2020hqc} are illustrated. {The y-axis represents the number of events per 0.1 MeV.} The blue and green curves are singles from the $\nu_x+^{13}{\rm C}$ NC and $\nu_x+e$ ES channels, respectively. The purple curve includes the $\nu_e+^{13}{\rm C}\rightarrow e^-+^{13}{\rm N}\,(\frac{3}{2}^{-})$ channel and the residual singles of the $\nu_e+^{13}{\rm C}\rightarrow e^-+^{13}{\rm N}\,(\frac{1}{2}^{-})$ channel after the coincidence cut. The red curve represents the single events from natural radioactivity and muon-induced unstable isotopes. The brown curve includes the $\Bar{\nu}_e+e$ ES and $\Bar{\nu}_x+^{13}{\rm C}$ NC channels from reactor antineutrinos. The black curve is the summation of all the components. The upper right insert plot is illustrated for the energy range between 3 and 5 MeV in the linear scale. Note that the discontinuities at 3 MeV and 5 MeV are due to the changes in the fiducial volume size.
    }\label{fig:specNC-ES}
\end{center}
\end{figure}

The typical signature for the NC event, $\nu_{x}+^{13}{\rm C}\rightarrow\nu_{x}+^{13}{\rm C}\,({3}/{2}^-;3.685\,{\rm MeV})$ is a mono-energetic $\gamma$ with the energy of $3.685\,{\rm MeV}$, {convoluted with the energy resolution of $\sigma_E/E=3\%/\sqrt{E \;({\rm MeV})}$ for the JUNO detector.}
The expected visible energy spectra of all single event sources for 10 years of data taking with the same energy-dependent fiducial volume cuts as in ~\cite{JUNO:2020hqc} are shown in Fig.~\ref{fig:specNC-ES}. The blue and green curves are singles from the $\nu_x+^{13}{\rm C}$ NC and $\nu_x+e$ ES channels, respectively. The purple curve includes the $\nu_e+^{13}{\rm C}\rightarrow e^-+^{13}{\rm N}\,({3}/{2}^{-};3.502\,{\rm MeV})$ channel and the residual singles of the $\nu_e+^{13}{\rm C}\rightarrow e^-+^{13}{\rm N}\,({1}/{2}^{-};{\rm gnd})$ channel after the coincidence cut.
{The anti-coincidence criterion successfully reduces residual singles from reactor antineutrino IBD interactions to a negligible level. These residuals are due to cases where prompt and delayed signals appear in the same 1
$\mu$s readout window.}
The red curve represents the single events from natural radioactivity and muon-induced unstable isotopes~\citep{JUNO:2020hqc}. The brown curve includes the $\Bar{\nu}_e+e$ ES and $\Bar{\nu}_x+^{13}{\rm C}$ NC channels from reactor antineutrinos. The NC events rate from reactor antineutrinos is less than 0.2\% of that from solar neutrinos. The black curve is the summation of all the components. Note that the discontinuities at 3 MeV and 5 MeV are caused by the energy-dependent fiducial volume cuts which are, from low to high energies, $R<$ 13 m for [2, 3] MeV, $R<$ 15 m for [3, 5] MeV, and $R<16.5\;{\rm m}$ for the energies large than 5 MeV.
The upper right insert plot is illustrated for the energy range between 3 to 5 MeV in the linear scale, where a clear peak from the solar neutrino NC channel can be seen above the continuous spectra from solar neutrino ES interactions and the other backgrounds, demonstrating the promising prospect for the observation of the NC channel at JUNO. After all the cuts the number of signal events in the NC channel is {738} for 10 years of data taking.


\subsection{$\nu_x+e$ Elastic Scattering Channel}

In this work, we follow exactly the same strategy as in ~\cite{JUNO:2020hqc} for the analysis of the $\nu_x+e$ ES channel, where energy spectra for the recoiled electrons as well as all the backgrounds have been shown in Fig.~\ref{fig:specNC-ES}. One should note that the upturn feature of the energy dependence of the solar neutrino survival probability is clearly visible in the electron energy spectrum.

\subsection{Day-Night Asymmetry}

The MSW effect can cause solar neutrino event rate variations as a function of the solar zenith angle when the neutrinos propagate through the Earth~\citep{Carlson:1986ui,Baltz:1987hn,Baltz:1988sv,Krastev:1988yu,Blennow:2003xw,Akhmedov:2004rq,deHolanda:2004fd,Liao:2007re,Long:2013ota}, and result in the day-night asymmetry of the solar neutrino observation, in which the signal rate in the night is higher than that in the day due to $\nu_e$ regeneration inside the Earth.

\begin{figure*}
\begin{center}
	\includegraphics[width=0.6\textwidth]{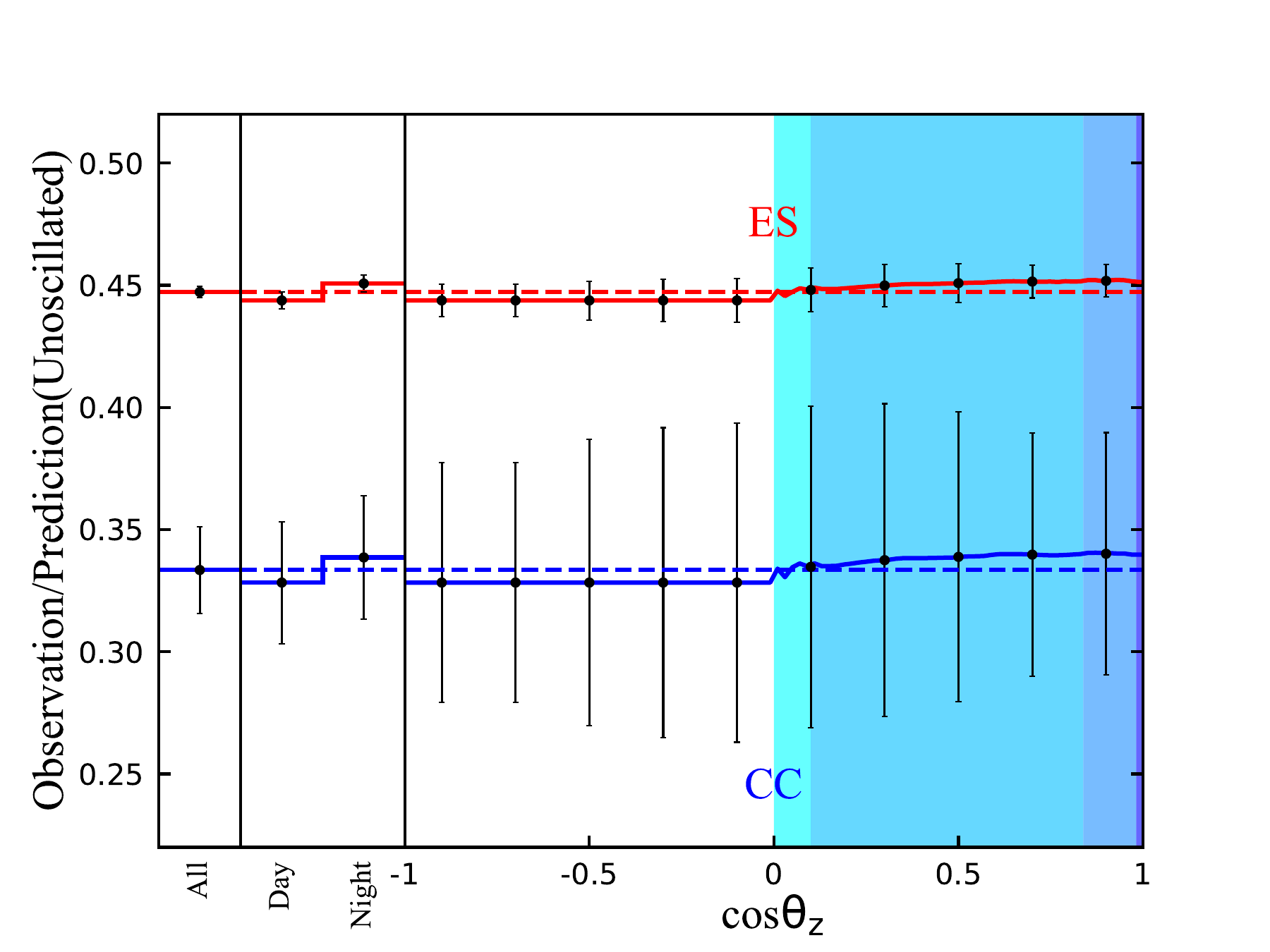}
    \caption{Ratios of the solar neutrino signal event rates with and without considering terrestrial matter effects as the function of the zenith angle for the ES (red sold line) and CC (blue solid line) channels. The dashed lines are shown for the average over the whole zenith angle range. The ratios for the daytime, nighttime and the day-night average are also shown for comparison. The blue shaded regions with different colors from the left to right are used to denote the zenith angle ranges passing through the crust, mantle and core of the Earth. Note that the signal rate in the night is higher than that in the day due to the $\nu_e$ regeneration through the Earth. 
    }
    \label{fig:DNA}
\end{center}
\end{figure*}

In this work, in addition to the visible energy spectra of the CC, NC and ES channels, we also consider the day-night asymmetry to constrain the neutrino oscillation parameters. The location of JUNO (i.e., 112$^{\circ}$31'05'' E and 22$^{\circ}$07'05'' N~\citep{JUNO:2021vlw}) is used in the day-night asymmetry calculations,
and the two dimensional visible energy and zenith angle spectra are employed. For illustration, we show in Fig.~\ref{fig:DNA} the ratios of solar neutrino signal event rates with and without considering the terrestrial matter effects as the function of the zenith angle $\theta_{z}$. The red and blue solid lines are for the ES and CC channels, respectively. In comparison, the dashed lines are shown for the respective averages over the whole zenith angle range.
The ratios of the day-night average ($R_{\rm A}$), the daytime ($R_{\rm D}$), and the nighttime ($R_{\rm N}$) are also illustrated with the first three bins. 
The error bars are quoted as the statistical uncertainties of the signal and backgrounds. The blue shaded regions with different colors from the left to right are used to denote the zenith angle ranges passing through the crust, mantle and core of the Earth respectively.
The day-night asymmetry, defined as $(R_{\rm D}-R_{\rm N})/R_{\rm A}$, is predicted to be $-3.1\%$ and $-1.6\%$ for the CC and ES channels respectively.
The energy ranges of the CC and ES channels are [5, 14] MeV and [2, 16] MeV respectively. Given that all the neutrino flavors can be detected through the NC channel, no day-night asymmetry exists in the NC detection. 
Note that the magnitude of the day-night asymmetry strongly depends on the value of $\Delta m^2_{21}$. If $\Delta m^2_{21}$ is decreased {from the KamLAND measurement $7.5\times 10^{-5}\;{\rm eV}^2$~\citep{KamLAND:2013rgu} to} $6.1\times 10^{-5}\;{\rm eV}^2$ of the global solar neutrino data~\citep{Esteban:2020cvm}, 
the absolute values of the day-night asymmetry are also increased to $-4.2\%$ and $-2.2\%$ for the CC and ES channels, respectively.

\section{Sensitivity Study}
\label{sec3}

In this section, we study the physical potential for the model independent measurement of $^8$B solar neutrinos using CC, NC, and ES channels. Based on the typical event signatures, the full solar neutrino data can be separated into the correlated and single event data sets. As discussed in the previous section, all the three interaction channels from $^8$B solar neutrinos would contribute to the single event data set, while the correlated data set includes events from both the CC channel and the accidental coincidence of the ES channel. 

In this analysis, we consider the following systematic uncertainties. First, the uncertainty of detection efficiency is estimated to be 2\%~\citep{JUNO:2020hqc}, which is fully correlated for the the signal and background components of each data set, but uncorrelated between the coincidence and single event data samples. Second,  
the current uncertainty of the $^{13}{\rm C}$ cross sections from the model calculation is at the level of several percents~\citep{Fukugita:1988hg,Suzuki:2012aa,Suzuki:2019cra}, but the precision could be reduced to 1\% or better with large-scale modern shell-model calculations~\citep{Barrett:2013nh}. Therefore the uncertainties for the $^{13}{\rm C}$ CC and NC interaction are taken as {1}\% for the current study. A 0.5\% cross section uncertainty is used for the ES channel~\citep{Tomalak:2019ibg}. Third, the shape uncertainty of $^8$B solar neutrinos is taken from ~\cite{Bahcall:1996qv,Bahcall:1997eg}, and the uncertainties for the radioactive and muon-induced backgrounds are the same as those in ~\cite{JUNO:2020hqc}, namely, 1\% for $^{238}$U, $^{232}$Th and $^{12}$B decays, 3\% for $^8$Li and $^6$He decays, and 10\% for $^{10}$C and $^{11}$Be decays. A 2\% uncertainty is used for the single event from the reactor antineutrino ES interaction.
In this work we treat the $^8$B solar neutrino flux as a free parameter since we are performing a model independent measurement. Only in the scenario of combining with the SNO flux measurement, an uncertainty of 3.8\% is used as an informative prior.

The standard Poisson-type $\chi^2$ method using the Asimov data set~\citep{ParticleDataGroup:2020ssz} is employed to estimate the sensitivity to measure the $^8$B solar neutrino flux and the oscillation parameters sin$^2\theta_{12}$ and $\Delta m^2_{21}$, where different pull parameters are included in the $\chi^2$ function to account for the systematic uncertainties described in this section. More technical details on the construction of the $\chi^2$ function are provided in the Appendix.
In order to identify the contribution of each interaction channel, we divide the whole data sets into the correlated events, the single events within [3.5, 4.1] MeV, and the single events outside [3.5, 4.1] MeV, which correspond to the ${\rm CC}$, ${\rm NC}$, and ${\rm ES}$ measurements respectively.

\begin{figure*}
\begin{center}
	\includegraphics[width=0.8\textwidth]{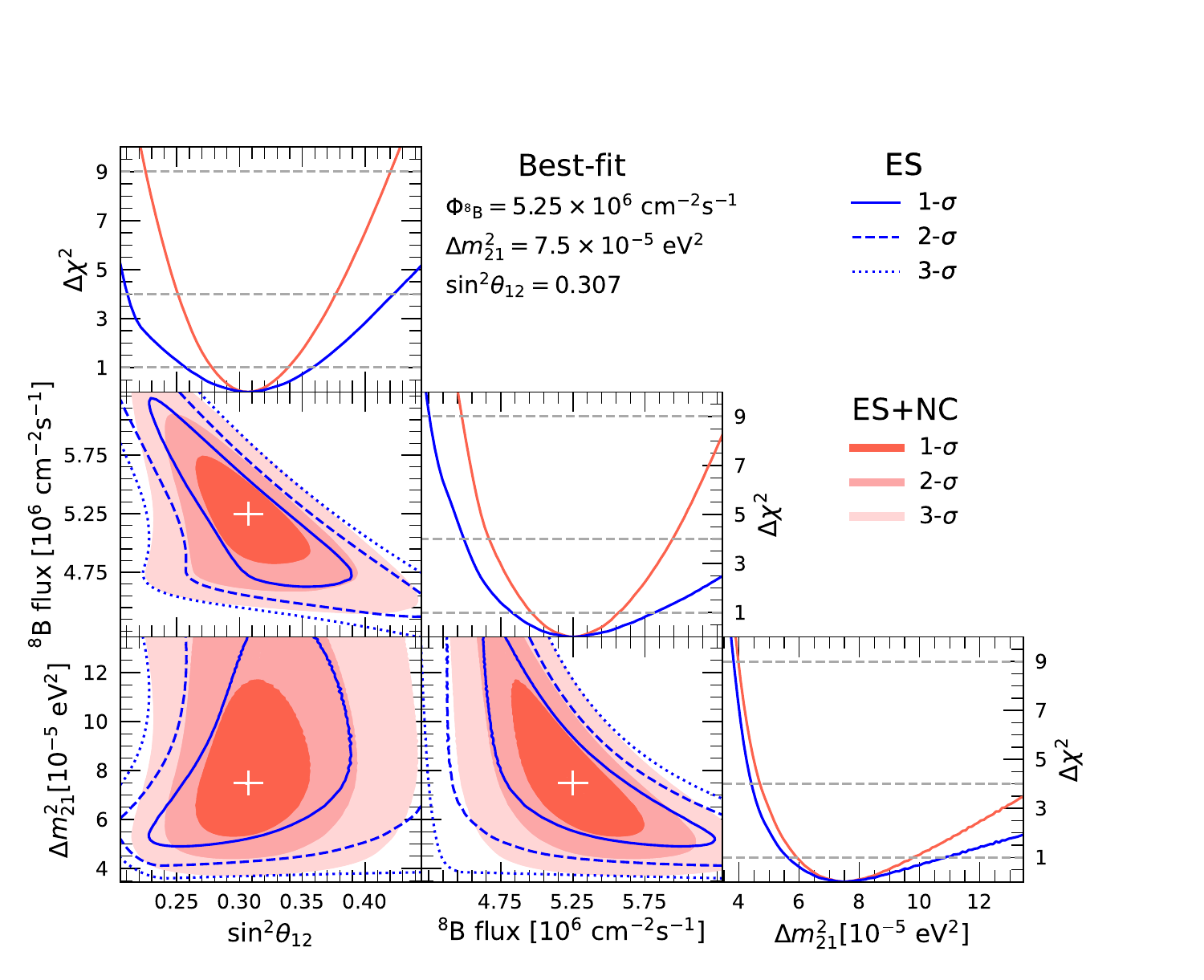}
    \caption{Comparison of the sensitivity on the $^{8}$B solar neutrino flux, sin$^2\theta_{12}$ and $\Delta m^2_{21}$ between the ES measurement (single events outside [3.5, 4.1] MeV) and the ES+NC measurement (all singles events). The 1$\sigma$ (68.3\%), 2$\sigma$ (95.5\%), and 3$\sigma$ (99.7\%) allowed regions are illustrated with blue lines and red shaded regions, respectively. The marginalized projections of these parameters are also shown.}
    \label{fig:2d1}
\end{center}
\end{figure*}

We illustrate in Figs~\ref{fig:2d1}-\ref{fig:2d3} the
two dimensional allowed ranges and the marginalized one dimensional curves on the sensitivity of the $^{8}$B neutrino flux, $\sin^2\theta_{12}$ and $\Delta m^2_{21}$, of which Fig.~\ref{fig:2d1} is
for the comparison of the ES and ES+NC measurements, Fig.~\ref{fig:2d2} for the comparison the ES+NC and ES+NC+CC measurements, and Fig.~\ref{fig:2d3} for the comparison of the JUNO and JUNO + SNO flux measurements. In addition, a summary of relative uncertainties on the $^8$B neutrino flux, $\sin^2\theta_{12}$ and $\Delta m^2_{21}$ from the model independent approach is provided in Fig.~\ref{fig:1dsummary}. Several important observations and comments are presented as follows.

\begin{itemize}
\item The NC measurement is accomplished based on the single events within [3.5, 4.1] MeV, where the background events are from the singles of ES and CC interactions of $^{8}$B solar neutrinos, together with the natural radioactivity and muon-induced unstable isotopes. The standard MSW effect of solar neutrino oscillations is used in the calculation of ES and CC interactions and the oscillation parameters sin$^2\theta_{12}$ and $\Delta m^2_{21}$ are marginalized. The $^{8}$B solar neutrino flux can be obtained with an accuracy of 10.6\% with the NC measurement, which is comparable to the level of 8.6\% from the NC measurement of the SNO Phase-III data~\citep{SNO:2011ajh}.

\begin{figure*}
\begin{center}
	\includegraphics[width=0.8\textwidth]{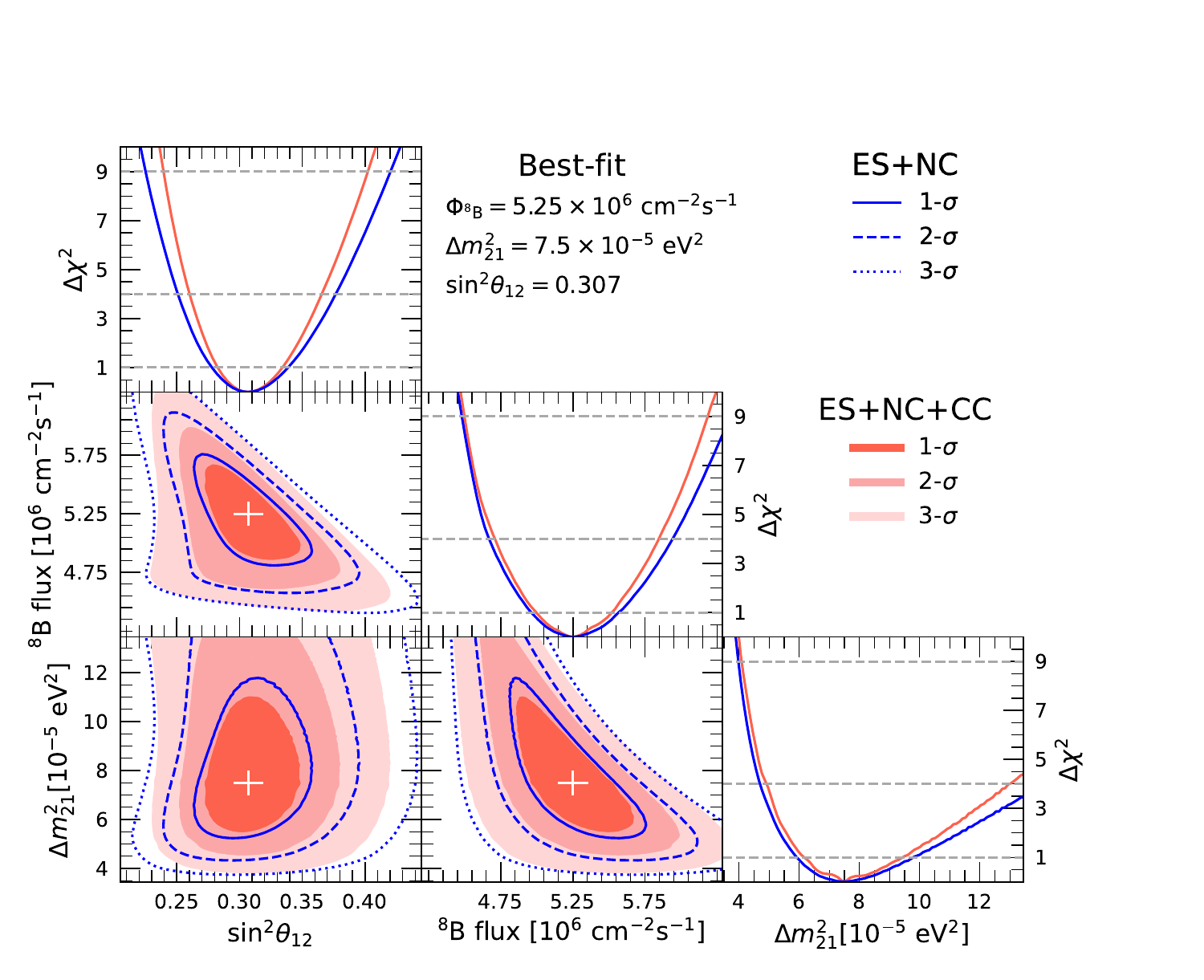}
    \caption{Same as Fig.~\ref{fig:2d1}, but for the comparison between the ES+NC measurement (all single events) and the ES+NC+CC measurement (both the single events and correlated events).}
    \label{fig:2d2}
\end{center}
\end{figure*}

\item The ES measurement is based on the single events outside the energy range of [3.5, 4.1] MeV, in which the dominant background is from the natural radioactivity and muon-induced unstable isotopes, which are summarized in Fig.~\ref{fig:specNC-ES} and more details can be found in ~\cite{JUNO:2020hqc}. In the model independent approach of the ES measurement, the $^{8}$B neutrino flux and two oscillation parameters sin$^2\theta_{12}$ and $\Delta m^2_{21}$ are simultaneously constrained, where the relative uncertainties are derived as $^{+11\%}_{-8\%}$, $^{+17\%}_{-17\%}$, and $^{+45\%}_{-25\%}$, respectively. The uncertainties of sin$^2\theta_{12}$ and $\Delta m^2_{21}$ are larger than those obtained in ~\cite{JUNO:2020hqc} by including the 3.8\% SNO flux measurement because of the strong correlation between the flux and oscillation parameters in the model independent approach. 
When adding the JUNO NC measurement, the accuracy of the $^{8}$B neutrino flux can be improved to the level of $^{+6.0\%}_{-5.5\%}$, and the uncertainties of sin$^2\theta_{12}$ and $\Delta m^2_{21}$ are also improved to $^{+10\%}_{-10\%}$, and $^{+31\%}_{-21\%}$ respectively.

\begin{figure*}
\begin{center}
	\includegraphics[width=0.8\textwidth]{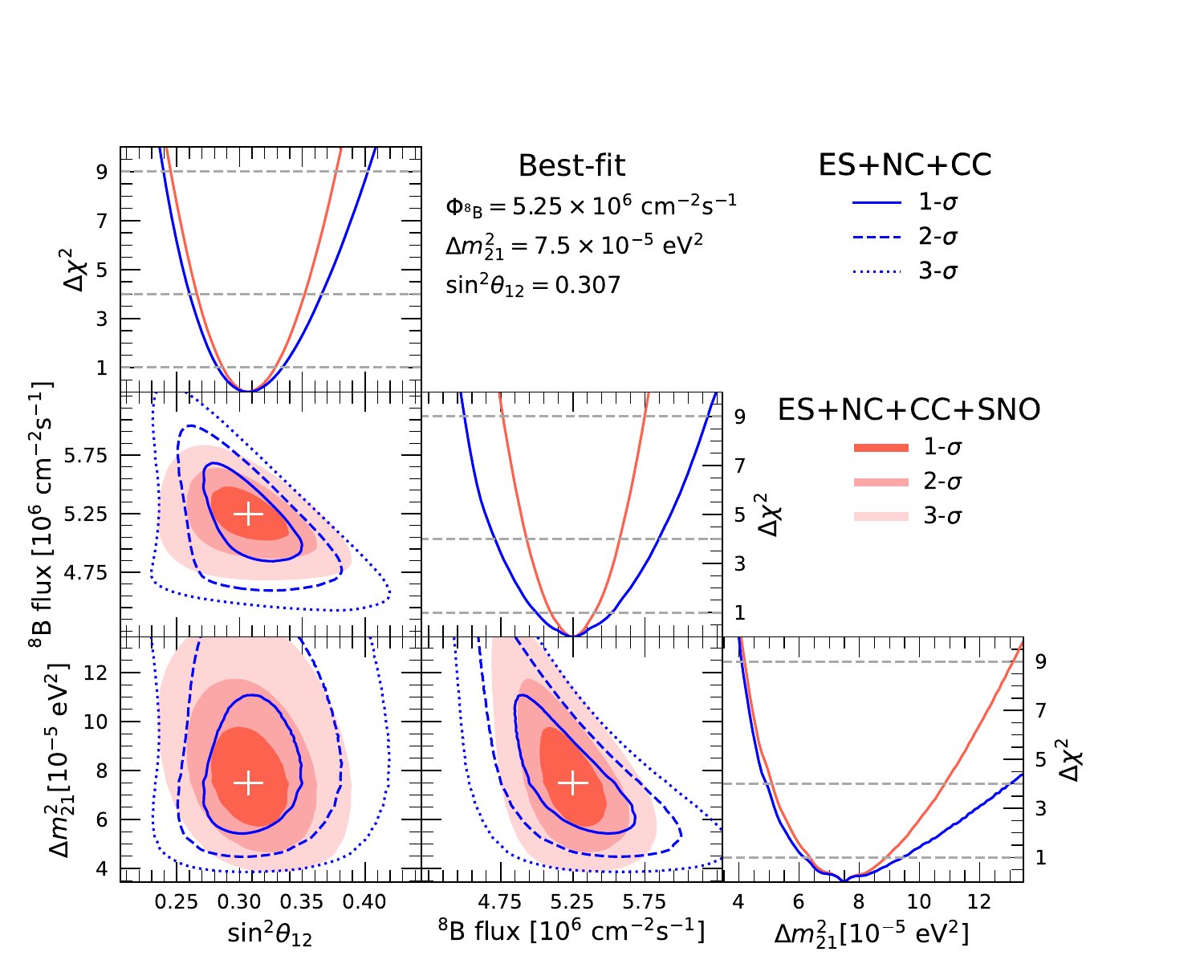}
    \caption{Same as Fig.~\ref{fig:2d1}, but for the comparison between the ES+NC+CC measurement of JUNO and the combined JUNO+SNO flux measurement.}
    \label{fig:2d3}
\end{center}
\end{figure*}

\item The CC measurement with the correlated events itself cannot simultaneously determine the $^{8}$B neutrino flux and oscillation parameters because of the high visible energy threshold. However, by combining the CC measurement with the single events of the NC+ES channels, it will help to break the correlation and possible degeneracy among different parameters, where the accuracy of the $^{8}$B neutrino flux can be further improved to 5\%, while those of $\sin^2\theta_{12}$ and $\Delta m^2_{21}$ are $^{+9\%}_{-8\%}$, and $^{+25\%}_{-17\%}$ respectively.

\item The expected 5\% precision of the $^{8}$B neutrino flux obtained with all three detection channels is much better than that of 11.6\% from the latest prediction of the SSM~\citep{Vinyoles:2016djt}. This will be the only model independent measurement after SNO~\citep{SNO:2011hxd}.
In addition, the uncertainties of $\sin^2\theta_{12}$ and $\Delta m^2_{21}$ from the $^{8}$B neutrino measurement at JUNO are at the levels of $^{+9\%}_{-8\%}$ and $^{+25\%}_{-17\%}$ respectively, which is comparable to the levels of $^{+5\%}_{-5\%}$, and $^{+20\%}_{-11\%}$ from the latest results of combined SK and SNO solar neutrino data~\citep{yasuhiro_nakajima_2020_4134680}. Considering that the reactor antineutrino measurement of JUNO will obtain sub-percent levels of $\sin^2\theta_{12}$ and $\Delta m^2_{21}$ in the near future~\citep{JUNO:2022mxj}, measurements of these parameters from future solar neutrino data would be important to test the CPT symmetry of fundamental physics and resolve the possible discrepancy between the neutrino and antineutrino oscillation channels.

\begin{figure*}
\begin{center}
	\includegraphics[width=0.8\textwidth]{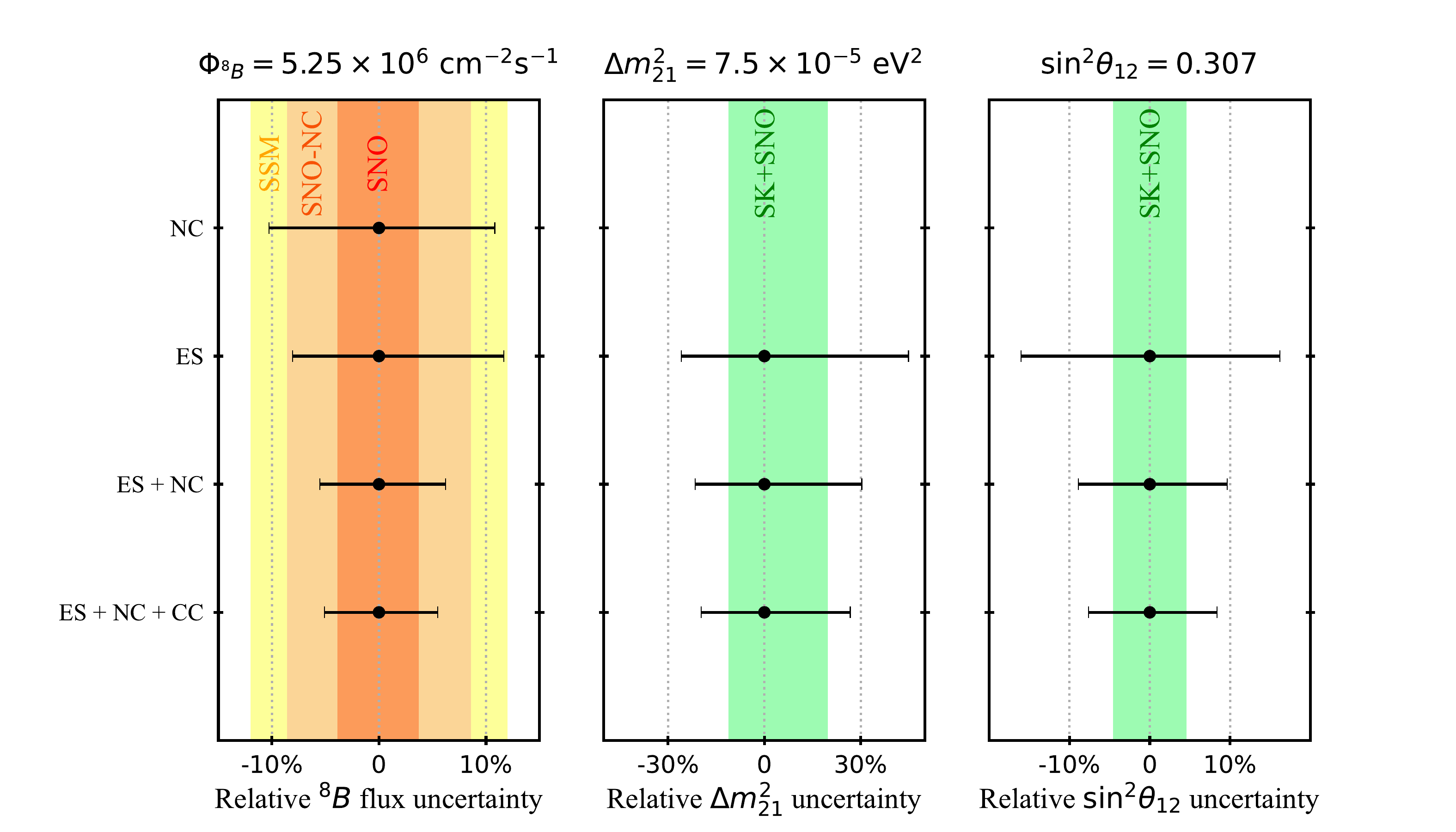}
    \caption{Relative uncertainties of the $^8$B solar neutrino flux (left panel), $\Delta m^2_{21}$ (middle panel), and $\sin^2\theta_{12}$ (right panel) from the model independent approach with different combinations of the data sets. The colored bands in the left panel are for the flux uncertainty from the SSM~\citep{Vinyoles:2016djt}, the NC measurement of the SNO Phase-III data (SNO-NC)~\citep{SNO:2011ajh}, and the combined SNO CC, NC and ES data (SNO)~\citep{SNO:2011hxd}. The green bands in the middle and right panels are the uncertainty of oscillation parameters from the combined SK and SNO solar neutrino data~\citep{yasuhiro_nakajima_2020_4134680}.}
    \label{fig:1dsummary}
\end{center}
\end{figure*}

\item Within the spirit of the model independent approach, one can also include the 3.8\% $^{8}$B neutrino flux measurement of SNO as an informative prior, where even better precision levels of the flux and oscillation parameters can be achieved. In this scenario, the expected accuracy of the $^{8}$B solar neutrino flux would reach the level of 3\%, and sin$^2\theta_{12}$ and $\Delta m^2_{21}$ can be constrained with the precision of $^{+7.5\%}_{-6.5\%}$, and $^{+19\%}_{-15\%}$ respectively. These measurements are comparable to those from the current global solar neutrino data and would provide unique information to the future solar neutrino program.

\item It is noteworthy that the {signal event statistics, detection efficiency} and cross section uncertainties are the most crucial factors that affect the detection potential of the CC and NC detection channels. If the cross section uncertainties are 10\%, instead of 1\% assumed in this work, the uncertainty of the $^{8}$B neutrino flux will become $^{+6\%}_{-6\%}$.

{\item In the CC detection channel, the observed energy of the prompt electron is directly related to the incoming neutrino energy, making it crucial to lower the prompt energy threshold to investigate the predicted increase in the solar neutrino survival probability at lower energies.
For this analysis, we set a conservative prompt energy threshold at 5 MeV to optimize the trade-off between the signal detection efficiencies and background contamination. 
Regarding the accidental background, radioactivity is the primary source of the prompt signal below 3.5 MeV, where stringent background control measures are essential, as outlined in ~\cite{JUNO:2021kxb}. Conversely, solar neutrino ES events become the leading prompt signal above 3.5 MeV. For the prompt energy range from 3.5 to 5 MeV, the cosmogenic correlated background is significantly higher than that in the region above 5 MeV, as depicted in Fig.~\ref{fig:CCspec}, while the signal efficiency is considerably lower between 3.5 and 5 MeV due to the multiplicity cut. Additional technical details in this regard will be reported elsewhere in the future.
}
\end{itemize}

\section{Concluding Remarks}
\label{sec4}

In this work we have studied the physics potential of detecting $^8$B solar neutrinos at JUNO, in a model independent manner by using the CC, NC and ES detection channels. Because of its largest-ever mass of $^{13}$C and the expected low background level, excellent signal-to-background ratios can be achieved.
Thus $^8$B solar neutrinos will be observable in all three interaction channels.

We have performed detailed evaluations of the background budgets and signal efficiencies of the CC, NC and ES channels at JUNO.
With optimized selection strategies, we find that the expected $^8$B neutrino rates of the CC and NC channels are $\mathcal{O}(100)$ interactions per year after the event selection.
It turns out that the {signal event statistics, detection efficiency and} cross section uncertainties are the most crucial factors that affect the detection potential of these two channels. 
We have carried out a combined analysis of both the coincidence and single events from all three detection channels, and shown that
the $^8$B solar neutrino flux, $\sin^2\theta_{12}$, and $\Delta m^2_{21}$ can be measured to $\pm5\%$, $^{+9\%}_{-8\%}$, and $^{+25\%}_{-17\%}$, respectively.
When combined with the SNO flux measurement, the world-best precision of 3\% can be achieved for the $^8$B neutrino flux.

In the history of solar neutrino experiments, the NC measurement is unique in decoupling the neutrino flux and oscillation parameters, and enabling the model independent approach of the solar neutrino program.
SNO has been the only solar neutrino experiment in the past to achieve this goal, and JUNO would be the second one.
In this work, we have demonstrated the feasibility of $^8$B solar neutrino measurements at JUNO, which, together with other large solar neutrino detectors~\citep{Capozzi:2018dat,Hyper-Kamiokande:2018ofw,Jinping:2016iiq}, will open a new era of solar neutrino observation and may uncover new directions for neutrino physics and solar physics.





 \medskip
\section*{Acknowledgements}
We are grateful for the ongoing cooperation from the China General Nuclear Power Group.
This work was supported by
the Chinese Academy of Sciences,
the National Key R\&D Program of China,
the CAS Center for Excellence in Particle Physics,
Wuyi University,
and the Tsung-Dao Lee Institute of Shanghai Jiao Tong University in China,
the Institut National de Physique Nucl\'eaire et de Physique de Particules (IN2P3) in France,
the Istituto Nazionale di Fisica Nucleare (INFN) in Italy,
the Italian-Chinese collaborative research program MAECI-NSFC,
the Fond de la Recherche Scientifique (F.R.S-FNRS) and FWO under the ``Excellence of Science – EOS” in Belgium,
the Conselho Nacional de Desenvolvimento Cient\'ifico e Tecnol\`ogico in Brazil,
the Agencia Nacional de Investigacion y Desarrollo and ANID - Millennium Science Initiative Program - ICN2019\_044 in Chile,
the Charles University Research Centre and the Ministry of Education, Youth, and Sports in Czech Republic,
the Deutsche Forschungsgemeinschaft (DFG), the Helmholtz Association, and the Cluster of Excellence PRISMA+ in Germany,
the Joint Institute of Nuclear Research (JINR) and Lomonosov Moscow State University in Russia,
the joint Russian Science Foundation (RSF) and National Natural Science Foundation of China (NSFC) research program,
the MOST and MOE in Taiwan,
the Chulalongkorn University and Suranaree University of Technology in Thailand,
University of California at Irvine and the National Science Foundation in USA.

\section*{Appendix}

In this appendix, we present the technical details of the sensitivity study employed in this work. 
A Poisson-type least squares function, denoted as $\chi^2$, is defined as follows,
\begin{eqnarray}
\chi^2 &=& \chi^2_{\rm stat}({\rm CC})+\chi^2_{\rm stat}({\rm NC})+\chi^2_{\rm stat}({\rm ES})+\chi^2_{\rm syst} \nonumber \\
&=&2\times\sum^{10}_{i=1}\Bigg[\sum^{90}_{j_{\rm C}=1}\left(N^{\rm C}_{\rm pre}(\theta^{i}_{z},E^{j_{\rm C}}_{\rm vis})-N^{\rm C}_{\rm obs}(\theta^{i}_{z},E^{j_{\rm C}}_{\rm vis})+N^{\rm C}_{\rm obs}(\theta^{i}_{z},E^{j_{\rm C}}_{\rm vis})\cdot {\rm log}\frac{N^{\rm C}_{\rm obs}(\theta^{i}_{z},E^{j_{\rm C}}_{\rm vis})}{N^{\rm C}_{\rm pre}(\theta^{i}_{z},E^{j_{\rm C}}_{\rm vis})}\right)\nonumber\\ 
&+&\sum^{140}_{j_{\rm S}=1}\left(N^{\rm S}_{\rm pre}(\theta^{i}_{z},E^{j_{\rm S}}_{\rm vis})-N^{\rm S}_{\rm obs}(\theta^{i}_{z},E^{j_{\rm S}}_{\rm vis})+N^{\rm S}_{\rm obs}(\theta^{i}_{z},E^{j_{\rm S}}_{\rm vis})\cdot {\rm log}\frac{N^{\rm S}_{\rm obs}(\theta^{i}_{z},E^{j_{\rm S}}_{\rm vis})}{N^{\rm S}_{\rm pre}(\theta^{i}_{z},E^{j_{\rm S}}_{\rm vis})}\right)\Bigg] \nonumber \\
&+&\left(\frac{\varepsilon_{X}^{\rm ES}}{\sigma_{X}^{\rm ES}}\right)^2+\left(\frac{\varepsilon_{X}^{\rm NC}}{\sigma_{X}^{\rm NC}}\right)^2+\left(\frac{\varepsilon_{X}^{\rm CC}}{\sigma_{X}^{\rm CC}}\right)^2+\sum_{k_{\rm C}}\left(\frac{\varepsilon^{k_{\rm C}}_{\rm B}}{\sigma^{k_{\rm C}}_{\rm B}}\right)^2+\sum_{k_{\rm S}}\left(\frac{\varepsilon^{k_{\rm S}}_{\rm B}}{\sigma^{k_{\rm S}}_{\rm B}}\right)^2 \nonumber \\
&+&\left(\frac{\varepsilon^{\rm C}_{\rm eff}}{\sigma^{\rm C}_{\rm eff}}\right)^2+\left(\frac{\varepsilon^{\rm S}_{\rm eff}}{\sigma^{\rm S}_{\rm eff}}\right)^2+\left(\varepsilon_{\rm s}\right)^2\,,
\label{eq:chi2}
\end{eqnarray}
where $\chi^2_{\rm stat}({\rm CC})$, $\chi^2_{\rm stat}({\rm NC})$, and $\chi^2_{\rm stat}({\rm ES})$ are statistical parts of the CC, NC and ES channels in the $\chi^2$ function, respectively. {These components are presented in the second and third rows of Eq.~(\ref{eq:chi2}). The index $j_{\rm C}$ ranges from 1 to 90 for the CC measurement, representing the energy range from 5 MeV to 14 MeV with an equal bin width of 0.1 MeV. For the NC measurement, $j_{\rm S}$ ranges from 16 to 21, while for the ES measurement, $j_{\rm S}$ spans from 1 to 15 and from 22 to 140 covering the energy range from 2 MeV to 16 MeV with an equal bin width of 0.1 MeV.}
{The predicted numbers of signal and background events,
$N^{\rm C}_{\rm pre}(\theta^{i}_{z},E^{j_{\rm C}}_{\rm vis})$ and $N^{\rm S}_{\rm pre}(\theta^{i}_{z},E^{j_{\rm S}}_{\rm vis})$
are calculated for the $i^{}$-th zenith angle bin and the $j_{\rm C}^{}$-th or $j_{\rm S}^{}$-th visible energy bin of the correlated and single event samples, respectively}
\begin{eqnarray}
N^{\rm C}_{\rm pre}(\theta^{i}_{z},E^{j_{\rm C}}_{\rm vis})&=&(1+\varepsilon_{\rm \rm eff}^{\rm C})S^{\rm CC}_{\rm pre}(\theta^{i}_{z},E^{j_{\rm C}}_{\rm vis})+\sum^{}_{k_{\rm C}}(1+\varepsilon_{\rm B}^{k_{\rm C}})B^{k_{\rm C}}_{\rm pre}(\theta^{i}_{z},E^{j_{\rm C}}_{\rm vis})\,,\\
N^{\rm S}_{\rm pre}(\theta^{i}_{z},E^{j_{\rm S}}_{\rm vis})&=&(1+\varepsilon_{\rm \rm eff}^{\rm S})[S^{\rm NC}_{\rm pre}(\theta^{i}_{z},E^{j_{\rm C}}_{\rm vis})+S^{\rm ES}_{\rm pre}(\theta^{i}_{z},E^{j_{\rm C}}_{\rm vis})]+\sum^{}_{k_{\rm S}}(1+\varepsilon_{\rm B}^{k_{\rm S}})B^{k_{\rm S}}_{\rm pre}(\theta^{i}_{z},E^{j_{\rm S}}_{\rm vis})\,,
\label{eq:spec}
\end{eqnarray}
{where $S^{\rm CC}_{\rm pre}$, $S^{\rm NC}_{\rm pre}$, and $S^{\rm ES}_{\rm pre}$ represent the two-dimensional spectra of the $^8$B neutrino signals in the CC, NC, and ES channels, respectively, incorporating the fiducial volume and signal efficiencies. The projections of these spectra onto the visible energy axis are depicted in Fig.~\ref{fig:CCspec} for the CC channel and Fig.~\ref{fig:specNC-ES} for the NC and ES channels. Meanwhile, $B^{k_{\rm C}}_{\rm pre}$ and $B^{k_{\rm S}}_{\rm pre}$ correspond to the background components in the correlated and single event samples, respectively, with their visible energy spectra illustrated in the same figures.} The calculations of the $^8$B neutrino signal spectra in the CC, NC, and ES channels are as follows:
{
\begin{eqnarray}
S^{\rm CC}_{\rm pre}(\theta^{i}_{z},E^{j_{\rm C}}_{\rm vis})& =& \Phi_{^8{\rm B}} \times \bigg[(1+\varepsilon_{\rm s}\delta^{\rm S}_{E_{\nu}})S_{^8{\rm B}}(E_{\nu}) \times P_{ee}(\theta_{12}, \Delta m^2_{21}, E_{\nu}, \theta^{i}_{z})\nonumber\\
&&\times (1+\varepsilon_{X}^{\rm CC})\otimes\sigma_{\rm CC}(E_{\nu},E_{e}) \otimes M_{\rm }(E_{e}, E^{j_{\rm C}}_{\rm vis})\bigg]\,,\\
S^{\rm NC}_{\rm pre}(\theta^{i}_{z},E^{j_{\rm C}}_{\rm vis})& =& \Phi_{^8{\rm B}} \times \bigg[(1+\varepsilon_{\rm s}\delta^{\rm S}_{E_{\nu}})S_{^8{\rm B}}(E_{\nu}) 
\nonumber\\
&&\times (1+\varepsilon_{X}^{\rm NC})\otimes\sigma_{\rm NC}(E_{\nu},E_{\gamma}) \otimes M_{\rm }(E_{\gamma}, E^{j_{\rm S}}_{\rm vis})\bigg]\,,\\
S^{\rm ES}_{\rm pre}(\theta^{i}_{z},E^{j_{\rm C}}_{\rm vis})& =& \Phi_{^8{\rm B}} \times \Bigg\{(1+\varepsilon_{\rm s}\delta^{\rm S}_{E_{\nu}})S_{^8{\rm B}}(E_{\nu}) \times \Sigma_{\alpha} \bigg[P_{e\alpha}(\theta_{12}, \Delta m^2_{21}, E_{\nu}, \theta^{i}_{z})\nonumber\\
&&\times (1+\varepsilon_{X}^{\rm ES})\otimes\sigma^{\nu_{\alpha}}_{\rm ES}(E_{\nu},E_{e})\bigg] \otimes M_{\rm }(E_{e}, E^{j_{\rm C}}_{\rm vis})\Bigg\}\,.
\label{eq:signal}
\end{eqnarray}}
{
The $^8$B neutrino signal spectra for 
the CC, NC, and ES channels are calculated by multiplying the $^8$B neutrino spectrum $S_{^8{\rm B}}$ with the neutrino oscillation probability $P_{e\alpha}$ (where $\alpha$ equals $e$ or $\mu+\tau$), and then convolving the resulting product with the differential interaction cross sections (namely, $\sigma_{\rm CC}$, $\sigma_{\rm NC}$, and $\sigma^{\nu_{\alpha}}_{\rm ES}$)  as well as with the detector response matrix $M_{\rm }$. The neutrino oscillation probability $P_{e\alpha}$ includes both the standard MSW flavor conversion and terrestrial matter effects, and is a function of the neutrino energy $E_{\nu}$ and the zenith angle $\theta^{}_{z}$, calculated within the three-neutrino oscillation framework. The detector response matrix $M_{\rm }$ accounts for the effects of energy resolution and energy non-linearity, as described in ~\cite{JUNO:2020hqc}.
The observed spectra $N^{\rm C}_{\rm obs}(\theta^{i}_{z},E^{j_{\rm C}}_{\rm vis})$ and $N^{\rm S}_{\rm obs}(\theta^{i}_{z},E^{j_{\rm S}}_{\rm vis})$ are obtained from the corresponding predicted spectra by applying the true values of the $^8$B neutrino flux $\Phi_{^8{\rm B}}$, oscillation parameters $\sin^2\theta_{12}$, and $\Delta m^2_{21}$, and assuming negligible contributions from nuisance parameters. Note that, as discussed in Sec.~\ref{sec2}, the $^8$B solar neutrino interactions may also contribute to the background components $B^{k_{\rm C}}_{\rm pre}$ (e.g., the green line in Fig.~\ref{fig:CCspec}, the purple line in Fig.~\ref{fig:specNC-ES}), In such instances, all correlations between the signal and background components are accounted for in the $\chi^2$ function.}

\begin{table}
\begin{center}
\footnotesize
\renewcommand\arraystretch{1.3}
 \caption{\label{tab:sysFit} Description the nuisance parameters and the associated uncertainties in the $\chi^2$ function.}
         \vspace{0.5cm}
	\begin{tabular}{c|c|c}
        \hline
    Sys. & Description for the pull term & Uncertainty \\ \hline
    $\varepsilon_{X}^{\rm ES}$, $\varepsilon_{X}^{\rm NC}$, $\varepsilon_{X}^{\rm CC}$ & Cross section for the CC, NC, ES channels & 1\%, 1\%, 0.5\% \\
    $\varepsilon^{\rm C}_{\rm eff}$, $\varepsilon^{\rm S}_{\rm eff}$  & Detector efficiency & 2\%~\citep{JUNO:2020hqc}  \\
    $\varepsilon_{\rm B}^{k_{\rm C}}$, $\varepsilon_{\rm B}^{k_{\rm S}}$ & Rate for the ${k_{\rm C}}$-th or ${k_{\rm S}}$-th background component & 1\%-10\%, same as ~\cite{JUNO:2020hqc} \\
    $\varepsilon_{\rm s}$ & $^8$B neutrino energy spectrum & ~\cite{Bahcall:1996qv,Bahcall:1997eg}  \\
    \hline
    \end{tabular}
\end{center}
\end{table}
{The nuisance parameters $\varepsilon^{m}_X$ ($m$=CC, NC, ES), $\varepsilon^{k}_{\rm B}$, $\varepsilon^{n}_{\rm eff}$ ($n$=C, S) account for systematic uncertainties associated with the cross section, the backgrounds, and the detection efficiency, respectively, as discussed in the manuscript. The parameter $\delta^{\rm S}_{E_{\nu}}$ represents the 1$\sigma$ fractional variation of the $^8$B neutrino energy spectrum, as detailed in ~\cite{Bahcall:1996qv,Bahcall:1997eg}, while $\varepsilon_{\rm s}$ denotes the magnitude of the $^8$B neutrino spectral uncertainty. A summary of the nuisance parameters and their corresponding uncertainties within the $\chi^2$ function is summarized in Table~\ref{tab:sysFit}. For the sensitivity study that produced the results shown from Fig.~\ref{fig:2d1} to Fig.~\ref{fig:1dsummary}, we selected data sets from one or a combination of the CC, NC, and ES measurements. We then activated the relevant nuisance parameters to account for systematic uncertainties in the corresponding  $\chi^2$ function. During the calculation of the allowed regions for each analysis, the displayed parameters (one or two of the fitting parameters $\Phi_{^8{\rm B}}$, $\theta_{12}$, and $\Delta m^2_{21}$) were fitted, while all other physical and nuisance parameters were marginalized. The critical values of $\Delta\chi^2$ for various confidence levels are sourced from ~\cite{ParticleDataGroup:2020ssz}.}

\bibliographystyle{aasjournal} 
\bibliography{reference}

\end{document}